\documentstyle[12pt,draft,epsf,rotate]{article}

\newcommand \be  {\begin{equation}}
\newcommand \ee  {\end{equation}}
\newcommand \bea {\begin{eqnarray} \nonumber }
\newcommand \eea {\end{eqnarray}}

\newcommand \gb  {\beta}

\newcommand \gr  {\rho}

\newcommand \gD  {\Delta}

\begin{document}

\title{How (Super) Rough is the Glassy Phase of a
Crystalline Surface with a Disordered  Substrate?}
\author{Enzo Marinari$^{(a)}$, Remi Monasson$^{(b)}$
        and Juan J. Ruiz-Lorenzo$^{(b)}$\\[0.5em]
  {\small (a): Dipartimento di Fisica and Infn, Universit\`a di Cagliari}\\
  {\small \ \  Via Ospedale 72, 09100 Cagliari (Italy)}\\
  {\small (b): Dipartimento di Fisica and Infn, Universit\`a di Roma
    {\em La Sapienza}}\\
  {\small \ \  P. A. Moro 2, 00185 Roma (Italy)}\\[0.3em]
   {\small \tt marinari@ca.infn.it}\\
   {\small \tt monasson@roma1.infn.it}\\
   {\small \tt ruiz@chimera.roma1.infn.it}\\[0.5em]}
\date{March 7, 1995}
\maketitle

\begin{abstract}
We discuss the behavior of a crystalline surface with a disordered
substrate. We focus on the possible existence of a {\em super-rough}
glassy phase, with height-height correlation functions which vary as
the square logarithm of the distance. With numerical simulations we
establish the presence of such a behavior, that does not seem to be
connected to finite size effects. We comment on the variational
approach, and suggest that a more general extension of the
method could be needed to explain fully the behavior of the model.
\end{abstract}

\vfill

\begin{flushright}
  { \tt cond-mat/9503074 }
\end{flushright}
\newpage

\section{\protect\label{S_INT} Introduction}

Recently the two letters \cite{CULSHA,BATHWA}(and a related comment,
\cite{BATHWA}) have stressed, by obtaining new numerical results, the
interest of a problem that can be described in the first instance as
the one of the surface of a crystal deposited on a disordered
substrate.

The model has at today a long story. Renormalization Group ideas
have been applied at first \cite{CAROST,TONDIV,TSASHA}, while
more recently the M\'ezard and Parisi \cite{MEZPAR} variational
approximation has lead to the drawing of a quite different picture
\cite{BOMEYE,GIALED,KORSHU}.

The relevant universality class describes indeed many and different
physical situations. The first one, that we have already quoted, is
the model of a crystal deposited on a disordered substrate. A second
one is a two dimensional array of flux lines with the magnetic field
parallel to the superconducting plane in presence of random pinning.
Close to the phase transition (whose existence is predicted by the
Renormalization Group and the by Variational Theory) the two models are
expected to have the same critical behavior. The universality class is
the one of the $2D$ Sine-Gordon model with random phases.

Let us define our system.  The dynamical variables of the model are
the integral displacements $d(x,y)$ of the surface from a disordered
bidimensional substrate. The variables $x$ and $y$ take integral
values going from $1$ to $L$.  The number of points of the lattice is
$S=L^2$.  The displacements $d(x,y)$ take positive, negative or zero
integral values.  The disordered substrate is characterized by
quenched random heights $\gr(x,y)$ in the range
$(-\frac{a}{2},+\frac{a}{2})$, where $a$ is the elementary step of the
surface columns (and will be one in the numerical simulations).
The total height
of the surface on the elementary $(x,y)$ square is

\be
  h(x,y) \equiv \ a\ d(x,y) + \gr(x,y)\ .
\ee
The Hamiltonian of the system is

\be
   H \equiv \frac{\kappa}{2} \sum_{x,y=0}^L
    \Bigl[
      \bigl( h(x,y)-h(x+1,y) \bigr)^2 + \bigl( h(x,y)-h(x,y+1) \bigr)^2
    \Bigr]\ ,
\ee
where in the numerical simulations
we will put the surface tension $\kappa$ equal
to two. The partition function will be defined as

\be
  Z_\gr \equiv \sum_{\{d(x,y)\}}e^{-\gb H}\ .
\ee
We will consider a quenched substrate, i.e. the free energy $F$ will
be defined as

\be
  F \equiv -\overline{\log Z_\gr}\ .
\ee
We will discuss  here mainly about the height-height correlation
function, which we define as

\be
  C(d) = \langle \bigl(  h(\vec{r}_0+\vec{d})
                       - h(\vec{r}_0) \bigr)^2  \rangle\ ,
  \protect\label{E_CD}
\ee
where we only take the $2$ dimensional vector $\vec{d}$ of the form $(d,0)$
or $(0,d)$, and by $\langle \cdot \rangle$ we denote collectively the
average over the different realizations of the noise, over the different
origins and the thermal average.

In the Gaussian model with integral variables and no disorder, the
surface is rough for $T>T_R$ \cite{CHUWEE}.  In the warm phase the
$C(d)$ of eq.  (\ref{E_CD}) behaves as $\log(d)$.  When $T<T_R$ the
surface becomes flat, glued to the ordered bulk.

When one considers the case of a disordered substrate the situation is far
less easy to analyze. The traditional approach to the problem is the one
based on Renormalization Group ideas, while only recently the variational
approximation approach by M\'ezard and Parisi \cite{MEZPAR} has been
applied to the problem. The results one obtains in the two approaches
have something in common. Both approaches find that there is a
transition at $T=T_R=\frac{\kappa}{\pi}$. In the high $T$ phase thermal
fluctuations make the quenched disorder irrelevant, and the systems
behaves as the pure model. Here correlations behave logarithmically, i.e.

\be
  C_{T>T_R}(d)\simeq \frac{T}{\kappa\pi} \log(d)\ .
  \protect\label{E_C1}
\ee
The differences come for $T<T_R$. In the Renormalization Group approach
\cite{CAROST,TONDIV,TSASHA}
one gets a new $\log^2 d$ dominant contribution. Here one finds that

\be
  C^{(RG)}_{T<T_R}(d)\simeq a_1 \log(d) + a_2 \log^2 d\ ,
\ee
where $a_1$ is non-universal, and $a_2$ is

\be
  a_2 \equiv \Bigl( \frac{T_R-T}{T_R} \Bigr)^2 \frac{2}{\pi^2}\ .
\ee
The presence of such a {\em super-rough} phase (where by super-rough
we imply a $\log^2 d$ behavior of the height-height correlation
functions) is indeed an interesting potential implication of the
presence of quenched disorder.  Such a behavior would imply that the
low $T$ phase is rougher than the high $T$ phase, that is quite
unusual. At high $T$ thermal fluctuations are able to carry the
surface away from the deep (but not deep enough) potential wells due
to the quenched disorder. So the roughening is the same than for the
pure model. At low $T$ the surface gets glued to the bulk. In the
ordered case this makes the surface smooth, since the bulk is
ordered. But in presence of the quenched disordered substrate this
effect does not smooth the surface, but on the contrary forces it to
follow a very rough potential landscape. This mechanism could force a
super-rough behavior.

The application of the variational approximation \cite{MEZPAR}
to this system \cite{BOMEYE,GIALED,KORSHU} does not lead to presence of a
$\log^2 d$ term, but to a behavior similar to the one of the high $T$
phase, with a slope of the logarithmic term which freezes at the critical
point

\be
  C^{(VAR)}_{T<T_R}(d)\simeq \frac{T_R}{\kappa\pi} \log(d)\ .
  \protect\label{E_C3}
\ee
We will try to argue in section (\ref{S_ANA}) that in some sense this is an
intrinsic limit of a too straightforward application of the variational
approximation (originally discussed for systems with continuous replica
symmetry breaking \cite{MEZPAR}) to systems with a single step broken
replica symmetry, and we will suggest that a more complex approach could
be needed in order to get a fair picture of this kind of systems.

A numerical analysis of references \cite{BATHWA,CULSHA} was making
indeed the mystery even greater. Systems which should belong to the
same universality class seem to show a very different
behavior. Reference \cite{BATHWA} was unable to detect any signature
of the glass transition when measuring static quantities in a
continuum random phase model, that, as we said, should belong to the
same universality class of our discrete model (but see the comment
\cite{BATHWA}). The authors of \cite{CULSHA} study the model we have
defined before, and seem to detect numerically a picture compatible
with the variational ansatz.  The approach suggested from Cule and
Shapir seemed to us interesting, and worse to be pursued further. It
has motivated us to run further simulations and more analysis of the
numerical data, and to look better in the theoretical problem of
selecting the correct analytic approach.

\section{\protect\label{S_NUM} The Numerical Simulations}

We have ran our numerical simulations on the APE parallel computer
\cite{APE}. Our code, all written in a high level language and very
elementary, was running at the $20\%$ of the theoretical maximal
speed.  The clear limit was the memory to the floating point unit
bandwidth, that in our way to write the problem was limiting us to the
$25\%$ of the theoretical efficiency.  It would be surely possible and
not very difficult to rewrite the code to obtain with an efficiency
close to $50\%$. Our code was running at a sustained performance
close to one Gflops on a APE {\em tube} (with a theoretical optimal
performance close to the $5$ Gflops).

Our program was truly parallel, in the sense each lattice was divided
among many processors. For example on a APE {\em tube}, which has $128$
processors arranged in a $3$ dimensional tubular shape of
$2\times 2\times 32$ we were running a single lattice on $4$ processors,
and we were running in parallel $32$ different random substrates in the
third processor direction. With this approach the smaller lattice we
could simulate is $4\times 4$. Our actual runs have all been using $L=64$
and $L=128$, simulating $256$ different substrate realizations and by
evolving two uncoupled replica's of the system in each random substrate
(with a total of $512$ systems). The average over the disorder was taken
over such $256$ realizations of the random quenched substrate.

We have started from a high value of $T$, running simulations for
decreasing $T$ values. For $L=128$ we have used temperatures of
0.9, 0.8, 0.7, 0.65, 0.6, 0.45 and 0.35, while for $L=64$ we have
used the values 1.0, 0.95, 0.9, 0.85, 0.8, 0.75, 0.7, 0.65, 0.6, 0.55, 0.50,
0.45, 0.40, 0.35, 0.30. At each $T$ value our run was starting from the
last configuration of the higher $T$ value. We have been very
conservative in requesting a long thermalization. At each $T$ value we
have added for $L=64$ $.5$ millions full Monte Carlo sweeps of the lattice
($.7$ millions for $L=128$), and then we have measured the correlation
functions $100$ times during $100,000$ further lattice sweeps. That
turned out to guarantee a good statistical determination of the
correlation functions. To check that in better detail we have chosen two
$T$ values, one in the warm phase and one in the cold phase, i.e. $T=.60$
and $T=.35$ for $L=64$. On this lattice, starting from the final
configurations, we have added first a series of $100,000$ more lattice
sweeps, and measured expectation values again. Then we have repeated the
procedure (all the measurements and the statistical analysis) by doubling
the added run (i.e. with $200,000$ added sweeps), and by doubling it
again (with $400,000$ added sweeps), and again (with $800,000$ added
sweeps). For both $T$ values all results were compatible, and no
transient effects were detected. The data points for the $L=128$ are
always very similar to the ones on the smaller lattice, in all our range
of temperatures. The dynamics was a simple Metropolis
Monte Carlo simulation.

Let us note that all our numerical data are fully compatible
(even if based on larger lattices and a better statistics), as far as
we have been able to check, with the data of reference \cite{CULSHA}.
What differs here is the analysis of the data,
and the fact that a more extensive data sample allows us to look in
better detail to the relevant quantities. We will detect here a small
effect, and the high statistics we have is crucial to be sure it is
significant. We stress the importance of comparing the full set of
correlation functions, at all distances, with the lattice form computed
on the same value of the lattice size $L$. We also believe it is
important to use the discrete form both for picking up the logarithmic
behavior and for picking up the super-rough behavior which is dominated
by a squared logarithm.

The lattice Gaussian propagator, which reproduces in the continuum
limit the logarithmic behavior, is

\begin{equation}
  P_L(d) \equiv \frac{1}{2 L^2} \sum_{n_1=1}^{L-1} \sum_{n_2=0}^{L-1}
  \frac{1-\cos(\frac{2 \pi d n_1}{L})}
  {2-\cos(\frac{2 \pi n_1}{L})-\cos(\frac{2 \pi n_2}{L})}\ .
  \protect\label{E_GAURET}
\end{equation}
As we have already stressed we also need a lattice transcription of the
squared logarithmic term. It is natural to take

\be
  P_L^{(2)}(d) \equiv P_L(d)^2\ .
  \protect\label{E_GAURE2}
\ee
These are indeed the terms we have used to fit our numerical data and to
try to distinguish a logarithmic behavior from a different asymptotic law.

In the following we will be comparing two possible behaviors of the
correlation function $C(d)$. One is the Gaussian scaling

\be
  C(d) = a_0 + a_1 P_L(d)\ ,
  \protect\label{E_FITGAU}
\ee
while the second includes a quadratic term, i.e.

\be
  C(d) = a_0 + a_1 P_L(d) + a_2 P_L^{(2)}\ .
  \protect\label{E_FITNON}
\ee
In the case of an high
temperature $T$, in the high $T$ phase,
the Gaussian fit to the correct lattice propagator
is very successful, and the non-Gaussian best fit
gives a non-linear contribution
compatible with zero. In this region we do not encounter any problem.

\begin{figure}
\begin{center}
\leavevmode
\epsfysize=350pt\rotate[r]{\hbox{\epsfbox[107 3 586 757]{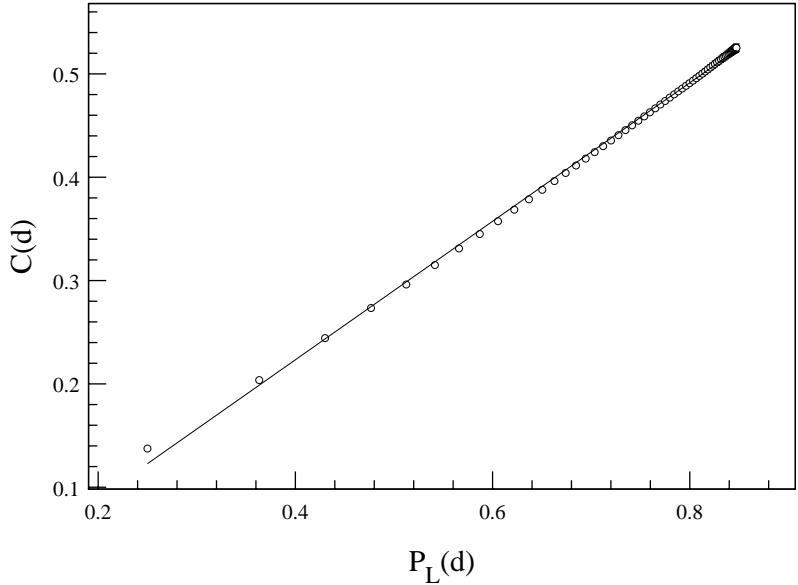}}}
\end{center}
  \protect\caption[1]{
    The measured correlation function $C(d)$
versus the lattice Gaussian propagator, at $T=0.45$ on a lattice of
size $L=128$. The straight line is our best fit to a linear behavior,
by using all distance data points.
    \protect\label{F_CORR}
  }
\end{figure}

We will discuss in the following the low $T$ region, and we will use
as an example the temperature $T=0.45$.
In fig. (\ref{F_CORR}) we plot the measured correlation function $C(d)$
versus the lattice Gaussian propagator, at $T=0.45$ on a lattice of
size $L=128$. A linear fit looks at this level quite satisfactory. The
discrepancy at low distances is not necessarily worrying, since we
expect short distance modifications to the asymptotic behavior.
We note for future comparison that the best fit gives here

\be
  C^{(all\ points)}_{(linear)}(d) = -0.02 + 0.62 P_L(d)\ ,
\ee
by using here all data points in the fit.  Let us note now that in
this fit and in all the following but for the quadratic one,
eq. (\ref{E_QUAFIT}), the errors (which we have estimated by a
jack-knife approach) are very small, of the order or smaller than one
percent.  All the best fits have been found be exact minimization of
the $\chi^2$ function, since in all cases it is quadratic in the
parameters.  The estimated linear coefficient is exactly the one one
finds in the variational approach (since the lattice propagator is
equal, in the continuum limit, to $(\mbox{const}\ +\
\frac{1}{2\pi}\log(r))$). A quadratic fit works here very well, but
since it has one more parameter than the linear fit let us ignore this
fact for a moment.  The quadratic fit of eq. (\ref{E_QUAFIT}), with
$3$ free parameters and done discarding $10$ distance points, has on
the contrary a very large error, but we report it for the indications
it gives about the reliability of the value we quote for the quadratic
coefficient (see the following discussion). Fitting including distance
points starting for example from $d=4$ would give a accurate
determination of all parameters.

\begin{figure}
\begin{center}
\leavevmode
\epsfysize=350pt\rotate[r]{\hbox{\epsfbox[107 3 586 757]{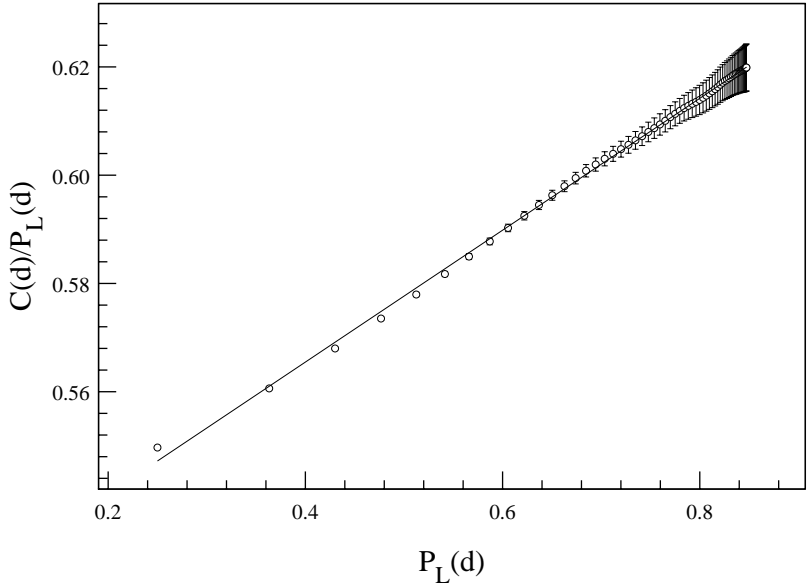}}}
\end{center}
  \protect\caption[1]{
    The measured correlation function $C(d)$ divided by the lattice
Gaussian propagator
versus the lattice Gaussian propagator, at $T=0.45$ on a lattice of
size $L=128$. The straight line is our best fit to a linear behavior,
by using all distance data points.
    \protect\label{F_CDIV}
  }
\end{figure}

As a next step we plot
in fig. (\ref{F_CDIV}) $C(d)$ divided by the lattice
propagator $P_L(d)$ as a function of $P_L(d)$. Linearity of this
quantity as a function of $P_L(d)$ implies the presence of a $\log^2$
term in the $C(d)$. The effect is very clear, and the evidence for the
presence of such a contribution unambiguous. We find here that

\be
  \Bigl(\frac{C(d)}{P_L(d)}\Bigr)^{(all\ points)}_{(linear)}
  = 0.52 + 0.12 P_L(d)\ ,
\ee
again by using data points from all distances.

\begin{figure}
\begin{center}
\leavevmode
\epsfysize=450pt\rotate[r]{\hbox{\epsfbox[107 3 586 757]{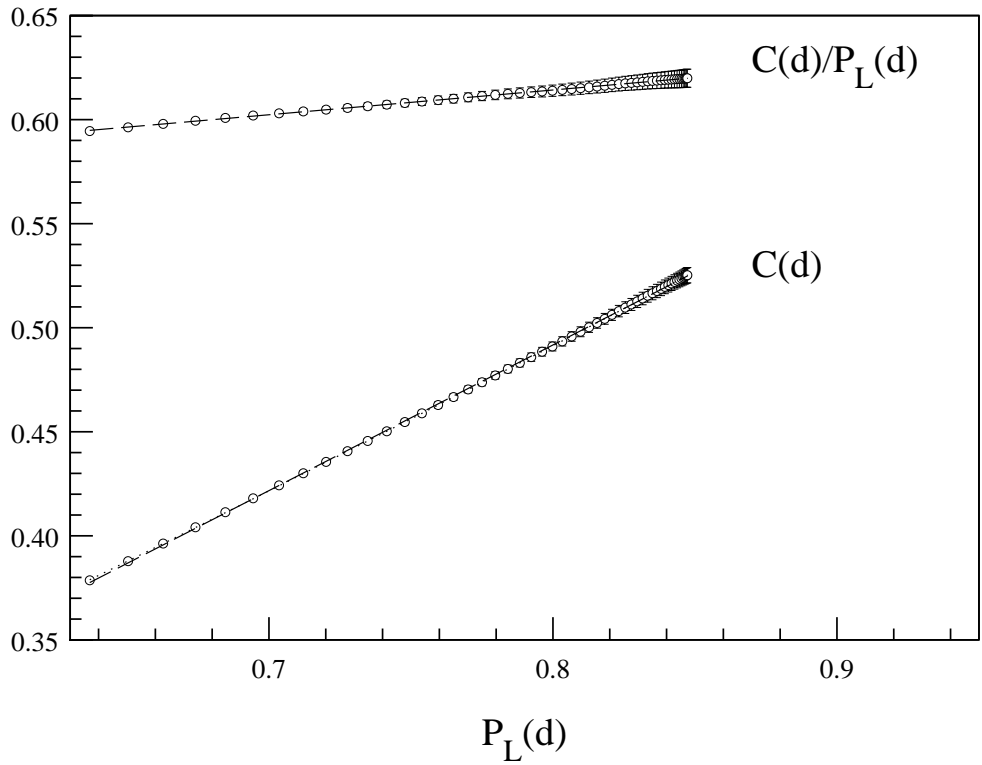}}}
\end{center}
  \protect\caption[1]{
    The measured correlation function $C(d)$ and
the measured correlation function divided by the lattice
Gaussian propagator
versus the lattice Gaussian propagator, at $T=0.45$ on a lattice of
size $L=128$, after discarding the first ten distance points.
The straight continuous lines are our best fits to a linear behavior,
while the dotted line is the bets fit to a quadratic behavior of the
function $C(d)$.
    \protect\label{F_CTEN}
  }
\end{figure}

The only concern we are left with is that in the previous analysis we
have used all distance points, while we are trying to resolve a long
distance behavior. We have to be careful not to be mislead by short
distance artifacts, which could obscure the true long distance
behavior. In order to play safe, we plot in fig. (\ref{F_CTEN}) both
$C(d)$ and  $\frac{C(d)}{P_L(d)}$ as a function of $P_L(d)$ for
distances larger than $10$ lattice units, and our best fits done by
using only these distance points. In this case we fit $C(d)$ both to
the linear and to the quadratic form. We get

\be
  C^{(d>10)}_{(linear)}(d) = -0.06 + 0.69 P_L(d)\ ,
\ee
that is very similar to our previous fit, and

\be
  C^{(d>10)}_{(quadratic)}(d) = - 0.014 + 0.56 P_L(d) + 0.093 P_L^{(2)}(d)\ .
  \protect\label{E_QUAFIT}
\ee
The most remarkable result is for the ratio we get

\be
  \Bigl(\frac{C(d)}{P_L(d)}\Bigr)^{(d>10)}_{(linear)}
  = (0.52 \pm .01) + (0.12 \pm .02)  P_L(d)\ ,
\ee
with a very small $\chi^2$ and in complete agreement with what we got
by fitting all data points. Two comments about two remarkable facts
are in order. In first the linear coefficient of the best fit for the
ratio is equal to the quadratic coefficient of the best fit for
$C(d)$, and the constant coefficient in the ratio fit is equal to the
linear coefficient of the quadratic fit to $C(d)$. In second
discarding ten short distance points does not change the results for
the linear and the quadratic contribution. We are definitely not
looking at a short distance effect.  Let us also notice that the
reader could think that since the error on the different points of the
divided $C(d)$ of fig. (\ref{F_CTEN}) are quite large the slope has to
be compatible with zero. This is not true since the data points
(different correlation functions for different $d$ values) are highly
correlated, and the error over the slope has to be estimated
directly. We have presented evidence that the value of the slope is
non-zero.

\begin{figure}
\begin{center}
\leavevmode
\epsfysize=350pt\rotate[r]{\hbox{\epsfbox[107 3 586 757]{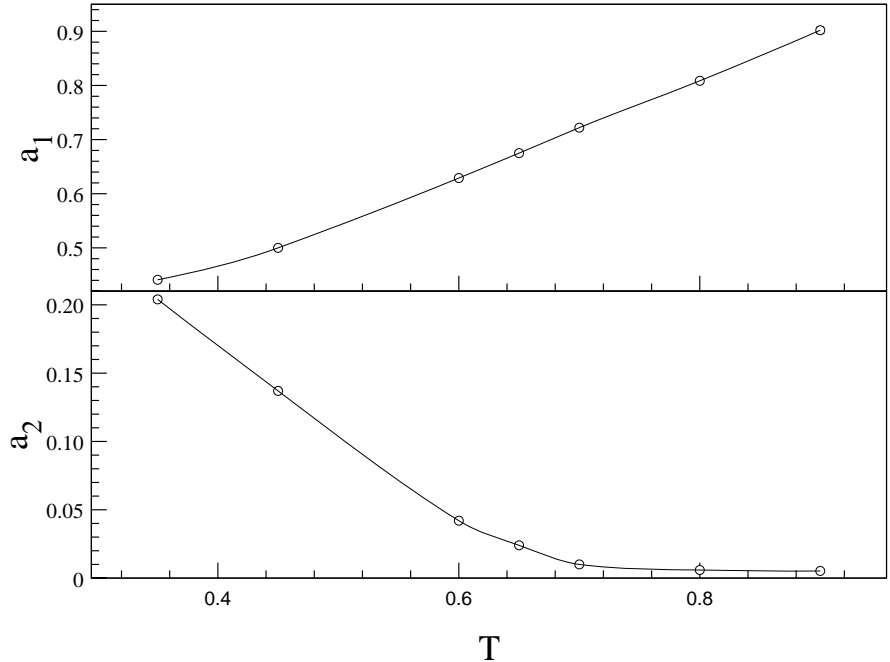}}}
\end{center}
  \protect\caption[1]{
  The coefficients $a_1$ and $a_2$ from our best fits
  to the form (\ref{E_FITNON}) versus the temperature $T$ in
  all the temperature range we have explored. Here $L=128$.
    \protect\label{F_COEF}
  }
\end{figure}

The fit to the form (\ref{E_FITNON}) gives a very good result both in
the warm phase (where it coincides with the gaussian fit) and in the
cold phase.  The presence of a lattice term corresponding to a
continuum $\log^2 d$ behavior accounts very well for our numerical
data. In fig.  (\ref{F_COEF}) we show the coefficients $a_1$ and $a_2$
from our best fits in all the temperature range we have explored (we
use here all distance points). The continuous lines are
only a visual aid, smoothly joining the numerical data points.  The
coefficient of the non-linear term $P_L^{(2)}$ becomes sizeably
different from zero exactly close to the critical temperature
$T_c=\frac{2}{\pi}$. The effect is quite clear and convincing.  The
coefficient $a_1$ is not the one of the $\log(d)$ term in the
continuum limit, that is renormalized by a contribution coming from
the $P_L^{(2)}(d)$ term.  We find that the coefficient of the
continuum $\log^2$ term is of the order of $5$ times smaller than the
universal value we would expect from the RG computations. This is a
fact that will have to be understood in better detail. The linear
dependence of $a_1$ over $T$ in the high $T$ phase, where $a_2=0$, is
very clear.

\begin{figure}
\begin{center}
\leavevmode
\epsfysize=350pt\rotate[r]{\hbox{\epsfbox[107 3 586 757]{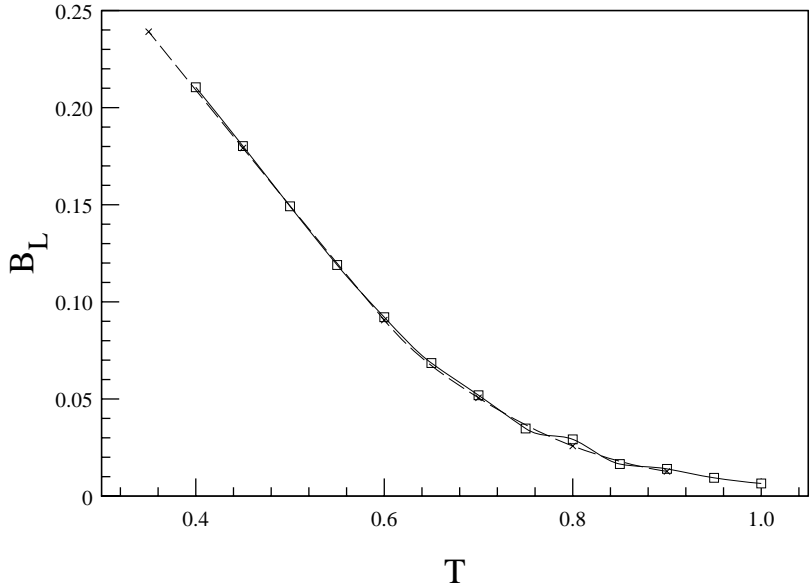}}}
\end{center}
  \protect\caption[1]{
    The Binder parameter $B_L^{(\gD)}$ as a function of $T$ for the two
    values of the lattice size, $L=64$ and $L=128$.
    \protect\label{F_BIND}
  }
\end{figure}

We believe that the previous evidence clearly shows that the ansatz of a
purely gaussian probability distribution does not explain the behavior of the
system for $T<T_R$, while the hypothesis of a super rough phase, with a
$\log^2 d$ behavior of the correlation functions, matches the numerical
findings very well. In order to gather more information about this glassy
phase we have looked at the probability distribution of

\be
  \gD\equiv h-h'\ ,
  \protect\label{E_DELTA}
\ee
where  $h'$ is a first neighbor of $h$. In order to monitor the shape of
the probability distribution we plot in fig. (\ref{F_BIND}) the related
Binder parameter defined as

\be
  \protect\label{E_BIND}
  B_L^{(\gD)}(T) \equiv \frac12\Bigl( 3 -
  \frac{\langle\gD^4\rangle}{\langle\gD^2\rangle^2}  \Bigr)\ .
\ee
$B_L$ is zero for a gaussian distribution, and $1$ for a
$\delta$-function. In our measurement it is very small in the warm phase,
calling again for a very Gaussian behavior. On the contrary in the low
$T$ phase we get a non trivial shape. Here $B_L$ is definitely non-zero,
non-one, and in our $T$ range does not seem to depend on $L$.
This shows again that in the cold phase there is a non-trivial behavior.
A value of $B_L$ which is non trivial (non $0$ and non $1$) and does not
depend on $L$ is reminiscent of a Kosterlitz-Thouless like situation.
Further analysis will be required to reach a better understanding of the
characteristic features of the low $T$ phase.

\section{\protect\label{S_ANA} Some Comments on the Variational Approach}

We have already argued that the numerical results presented above do
not coincide with the analytical predictions either of
the Renormalization Group or of the Gaussian Variational
approach. From the qualitative standpoint, the disagreement is even
stronger with the latter since it predicts that the asymptotic
correlation function grows only logarithmically with the
distance. Consequently, one has a right to wonder how much the
Gaussian Variational approach is trustworthy. Some general remarks on
this issue will be given in this section.

As argued by M\'ezard and Parisi in their original paper
\cite{MEZPAR}, the Gaussian Variational Theory (GVT) is exact for the
theory of an $N$-components field ${\vec \phi} ({\bf x})$ in the
limit $N \to \infty$ (while the model considered in this paper
corresponds to $N=1$). This may be easily understood by noticing that
the GVT coincides with the Hartree-Fock partial resummation of the
graphs due to the interaction potential between the replicas
\cite{MEZPAR}.  If the quenched potential $V({\bf x},{\vec \phi})$
seen at point ${\bf x}$ by the field ${\vec \phi}$ is itself a
Gaussian variable of zero mean and variance $\overline{ V({\bf
x},{\vec \phi}) V({\bf x}',{\vec \phi}')} = \delta ({\bf x} -{\bf x}'
) {\cal R} ({\vec \phi}-{\vec \phi}')$ that is, with a purely local
interaction, then the tadpole contribution to the self-energy $\sigma ^{ab}
(k)$ between two different replicas $a$ and $b$ will not
depend upon the momentum $k$ and will only result in renormalization of
the mass term. When the solution of the variational equations
turns out to be consistent with a continuously replica broken mass term
$\sigma (u)$ (where $0<u<1$ is the distance between the replicas),
this is not a serious limitation and non trivial exponents may be
found \cite{MEZPAR,GIALED,KORSHU},
related to the small $u$ behavior of $\sigma (u)$. On
the contrary, when the solution consists of a finite number of
breaking steps as in the present case, there is no such small $k$ -
small $u$ cross-over, and the full propagator is proportional to the
bare propagator at small $k$ (when $\sigma(0)=0$).  This explains why
the one-step Ansatz used in \cite{GIALED,KORSHU}
necessarily leads to a logarithmic
growth of the correlation function which corresponds to the free
behavior, the only non trivial prediction being the freezing of the
coefficient of proportionality under $T_R$ (see formulae (\ref{E_C1})
and (\ref{E_C3})).

An important example where the Gaussian Ansatz leads to erroneous
results is the Random Field Ising Model (RFIM). Recently, M\'ezard and
Young proposed a general method of adding momentum-dependent
contributions to the self-energy $\sigma ^{ab} (k)$ by considering a
self-consistent expansion in $1/N$ of the variational free-energy
\cite{MEZYOU}. Applying this method to the RFIM, they were able to
show that the new graphs at $O(1/N)$ improved the Gaussian
Ansatz and were sufficient to break the so-called dimensional
reduction coming from usual perturbation theory, and which of course
held at the Gaussian level. Such an approach could also be used to improve our
theoretical understanding of the model studied in this paper.  It
suffers however from a mathematical difficulty related to the absence of
solutions
$\sigma ^{ab} (k)$ with a finite number of steps of breaking, forcing
one to look for a fully broken mass $\sigma (k,u)$. So far, no solution
has been found in the case of the RFIM and one would probably have to
face the same difficulties for the Random Sine-Gordon Model.

Beyond the quantitative calculation of the critical exponents, an
important feature of the GVT is that it leads to a simple
determination of the phase diagram of the model studied here. In this
respect, the figures \ref{F_COEF} and \ref{F_BIND} seem to indicate
that the distribution of the height differences $\Delta$ defined in
(\ref{E_DELTA}) differs from a Gaussian even at temperatures higher
than the usual theoretical prediction $T_R={\kappa\over \pi}$. As the
two curves for the sizes $L=64$ and $L=128$ coincide quite well,
finite size effects can apparently not account for this
discrepancy. Some preliminary analytical results we have obtained
using the GVT above $T_R$ hint indeed at a possible dynamical
transition at a temperature $T_d$ ($T_d > T_R$) whose value depends on
the amount of disorder given by the variance of the quenched
displacement field $d(x,y)$ (see the introductive section). If this
would be so, there would already exist at the temperature $T_d$ an
exponentially large number (in $L^2$) of metastable states and the
system would only partially ``thermalize'' in these traps.  Both
numerical and analytical work is currently in progress to investigate
this important issue.

\section{\protect\label{S_CON} Conclusions}

The results we have obtained describe a very complex picture. A
super-rough behavior seems indeed to be there, implying that the GVT
does not account fully for the behavior of the model.  We are
discussing here very small effects, so we cannot exclude completely
that we are not looking at a transient behavior, but that does not
seem likely at all.  On the other side the coefficient of such a
non-linear term seems to be, far out of statistical and systematic
error, different from the one one obtains with a RG computation. Also,
a non-trivial Binder parameter looks non-trivial even in the beginning
of the to-be warm phase (that is maybe not the warm phase yet),
suggesting the presence of a complex scenario also for temperatures
$T$ of the order of $0.8$.

We have argued that indeed from the theoretical view-point we have
some understanding of what is happening. We  hope we will succeed to
deepen it in the near future.

\section*{Acknowledgments}
We acknowledge useful discussions with Marco Ferrero,
David Lancaster, Giorgio Parisi
and  Marc Potters. J.~J.~R.-L. is supported by a MEC grant (Spain).

\hfill

\end{document}